\documentclass[11pt,a4paper]{article}
\usepackage[top=0.85in,left=1in, right = 1in,footskip=0.75in]{geometry}

\usepackage[english]{babel}

\usepackage{multirow}
\usepackage{subfigure}
\usepackage{booktabs}
\usepackage{graphicx}
\usepackage{eurosym}
\usepackage{float}
\usepackage{color}
\usepackage{authblk}
\usepackage{setspace}
\usepackage{amsmath}
\usepackage{footnote}
\usepackage{amssymb}
\usepackage{url}
\usepackage{afterpage} 
\usepackage{soul}
\usepackage{wrapfig}
\usepackage{lscape}
\usepackage{rotating}
\usepackage{epstopdf}
\usepackage{placeins}
\usepackage{enumitem}
\usepackage{dcolumn}
\usepackage{hyperref}
\usepackage{tabulary,rotating, tabularx, siunitx}
\usepackage{setspace}
\usepackage{rotating}
\usepackage{tabularx}
\usepackage{scicite}
\usepackage{times}
\usepackage[displaymath, mathlines]{lineno}
\usepackage{changes}

\definechangesauthor[name=jannes, color=blue]{jannes}

\definechangesauthor[name=milan, color=red]{milan}

\definechangesauthor[name=koen, color=green]{koen}

\definechangesauthor[name=bruno, color=green]{bruno}

\title{\textsc{Network control by a constrained external agent as a continuous optimization problem}}

\author[a,c,d]{Jannes Nys\footnote{Corresponding author: Jannes.Nys@uantwerpen.be}}
\author[b,d]{Milan van den Heuvel} 
\author[b,d]{Koen Schoors}
\author[b]{Bruno Merlevede}

\affil[a]{Department of Computer Science, University of Antwerp - imec, IDLab, 2000 Antwerp, Belgium}
\affil[b]{Department of Economics, Ghent University, 9000 Ghent, Belgium}
\affil[c]{Department of Physics and Astronomy, Ghent University, 9000 Ghent, Belgium}
\affil[d]{Complex Systems Institute, Ghent University, 9000 Ghent, Belgium}

\date{\today}

\begin{document}
\newcommand{\iden}[1]{\mathbb{I}_{#1}}
\newcommand{\abs}[1]{\left|#1\right|}
\renewcommand{\vec}[1]{\mathbf{#1}}
\newcommand{\argmin}[1]{\underset{#1}{\text{argmin}}~}
\newcommand{\etal}{\textit{et al.}}
\newcommand{\Lp}[1]{\| #1 \|}
\maketitle

\newpage
\begin{abstract}
\noindent
Social science studies dealing with control in networks typically resort to heuristics or describing the static control distribution. Optimal policies, however, require interventions that optimize control over a socioeconomic network subject to real-world constraints. We integrate optimisation tools from deep-learning with network science into a framework that is able to optimize such interventions in real-world networks. We demonstrate the framework in the context of corporate control, where it allows to characterize the vulnerability of strategically important corporate networks to sensitive takeovers, an important contemporaneous policy challenge. The framework produces insights that are relevant for governing real-world socioeconomic networks, and opens up new research avenues for improving our understanding and control of such complex systems.

\end{abstract}

\clearpage

\section*{Introduction}
Networks are ubiquitous in modern society. The interconnectedness of networks has the potential to magnify the impact of a local event or intervention on the system as a whole. This has been shown across multiple fields, from cascades in financial systems~\cite{Roukny2013}, to  service adoption in social networks~\cite{Banerjee2013}. This raises the question how (a set of) nodes can gain influence or control over certain states in socioeconomic networks (e.g. decisions in companies, opinions of people). Insight into this process would improve our understanding of the stability of a plethora of systems (e.g. financial system, governmental system) and facilitate the design of optimal interventions to stabilize these systems or steer them in otherwise desirable directions. What is the optimal targeting strategy to maximize adoption of products or services in a socioeconomic network? How vulnerable is the control over strategic sectors to hostile takeovers? In this paper, we present a framework that reformulates these questions as a continuous optimization problem. This allows us to combine network science with state-of-the-art optimisation methods from deep learning to gain insights into the control of complex networks in a scalable and flexible manner.

In the social sciences, and especially the field of social influence analysis, control is viewed as a property that (instantly) flows through the system from node to node, like influence on product adoption in social networks, or corporate control through share holdings in ownership networks ~\cite{Banerjee2013, Rungi2017, Banerjee2020}. In social influence analysis, influence refers to the ability of an entity to change the probabilities of states of other entities in the network, (e.g. influencers that make their followers buy items). We use the term control in a broad sense to refer to influence over the distribution of states in the system, unless explicitly mentioned otherwise.
Information diffusion models model how information flows through the network~\cite{Li2018}. The literature uses algorithms on these networks to solve the social influence maximization (SIM) problem~\cite{Kempe2003}, defined as: \textit{the problem of finding the minimal subset of nodes, the seed nodes, in a socioeconomic network that could maximize the spread of influence}, which is NP-hard. This has caused the literature to mostly focus on solutions that are either approximate or use heuristics. These solutions fail to scale well for large networks~\cite{Banerjee2020}.

Another domain where the topic of network control has been studied is (structural) network controllability~\cite{Liu2011,Nepusz2012,Zhang2019}. Herein, the goal is to find a minimal set of nodes, the driver nodes, that have the potential to independently control all states in the network in finite time. The control is exerted by injecting an external signal into the driver nodes, which then propagates throughout the network, as dictated by a set of (linear) differential equations. In this literature the driver nodes are said to have control if they are capable of independently steering the states of all (target~\cite{Gao2014}) nodes, or edges~\cite{Pang2017} in the network to any desired final state. Control in this strict sense seems a useful concept for designing policy interventions in socioeconomic networks, and the method has been demonstrated on such networks in the past~\cite{Galbiati2013, egerstedt2011degrees, Heuvel2021}. As shown in van den Heuvel and Nys~\cite{Heuvel2021} however, controllability holds implicit communication assumptions which causes the results to have little meaning in socioeconomic networks. The assumption originates from the independence condition that is imposed on the control~\cite{Kalman1963} and entails that, for node controllability, nodes are only able to send a single message to all their neighbors, which in most socioeconomic contexts does not make sense. In the context of corporate control for instance, an often used example in both the node and edge controllability literature~\cite{Liu2011, Nepusz2012}, this would unrealistically imply that a company can only give a single instruction to all owned subsidiaries.

Taking inspiration from the external signal perspective in controllability (where an external node is connected to the driver nodes to steer the system), and the flow of control in social influence analysis (where models are utilised that describe the diffusion of control), we reformulate the question of optimal (static) control as the continuous optimisation of an objective function of a constrained external agent. This allows us to construct a framework that, given a differentiable control objective, leverages scalable automatic differentiation and gradient-based optimization strategies to compute the optimal (or most efficient) intervention an external agent can perform on the network to maximize its control. The function to optimize consists of two parts: a first term capturing the agent's control in a given network configuration, and a second term capturing the cost (e.g. effort, monetary value) to obtain this control. Optimizing this objective function then essentially boils down to maximizing the control at minimal cost. The external agent perspective does not limit the scope of our approach, as one can let the external agent coincide with an internal node and initialize it with its edges as a starting point of the optimisation.

Traditionally, techniques for optimizing an objective function can be classified into three categories 1) symbolic differentiation, 2) numerical differentiation through the finite difference approximation of gradients, and 3) automatic differentiation. However, in network science applications, symbolic differentiation becomes intractable, and numerical differentiation becomes unstable, due to the many variables and interacting components. In deep learning, the literature was faced with a similar challenge in the analysis of large systems~\cite{baydin2018automatic}, which has been tackled by automatic differentiation. While automatic differentiation has received little attention in network science, we show its usefulness by demonstrating that, under the condition that the objective function is continuous and differentiable almost everywhere~(the proof in \cite{lee2020correctness} holds for our method as well), automatic differentiation and gradient-based optimization strategies can be used to find an optimum in the context of network control. 

Given the many challenges to empirically measuring control in real-world socioeconomic networks~\cite{PENG201817} we will, as a first step, demonstrate our framework in the context of corporate ownership networks, where direct bilateral control can be defined through (partial) ownership of companies. 

The decrease in communication and information costs has made businesses more interdependent in global value chains~\cite{Baldwin2016}. The associated globalisation of the corporate control market has resulted in more complex and  internationalized corporate ownership. This has prompted several developed countries, including the EU and the US, to introduce or sharpen foreign investment screening mechanisms on the grounds of security or national interest~\cite{UNCTAD2019}. At the same time governing such networks has become more cumbersome. Establishing who has controlling power in a company has become a nontrivial task, further complicated by different accounting rules and company structures across countries~\cite{GarciaBernardo2017}. Individuals may control a holding company that acts as a store of ownership of other companies. Companies may control other companies indirectly through intermediary subsidiaries~\cite{Rungi2017}. A network approach has therefore become prerequisite for obtaining a deeper understanding of corporate control problems. To account for indirect control along longer paths, ownership has to be propagated and consolidated throughout the different ownership paths between firms. Vitali~\etal~\cite{vitali2011network} proposed an adequate algorithm that summarizes the control of a given company on the network (how many subsidiaries a company controls, how much monetary value they represent, etc.). Using our framework we leverage this bilateral control matrix between companies and the cost of control, here the cost of buying shares, to find the optimal intervention by an external agent (company, government, etc.) to maximize its control over (part) of the network.

The remainder of the paper is structured as follows. First, we present our framework to optimize interventions in a network to gain optimal control. We detail the construction and optimisation of the objective function and show how we can account for real-world constraints on interventions such as budgets and target groups. Second, we illustrate the power of our framework by applying it in the context of the corporate ownership network. We start by explaining the intervention problem in this context and applying our framework to a simple synthetic graph to gain insights into the algorithm. We end this section by using data from Orbis Europe on Great Britain's biotech research sector to show how the framework can be used to gain crucial insights into the vulnerability of this sector to hostile takeovers in different policy environments. 
In the final part of this paper we discuss new research opportunities that can be explored with our framework, potential future extensions, and  broader applicability.

\section*{Optimal Control Framework}

In most use cases, the focus is explicitly on obtaining control over the nodes in a network (rather than e.g. over the edges). We therefore introduce our methodology in this context. A more general introduction to our core method, and an extension for constrained optimization, is given in the Methods section. 

\subsection*{Node control optimization}

\begin{figure}[tb]
    \centering
    \includegraphics[width=0.6\textwidth]{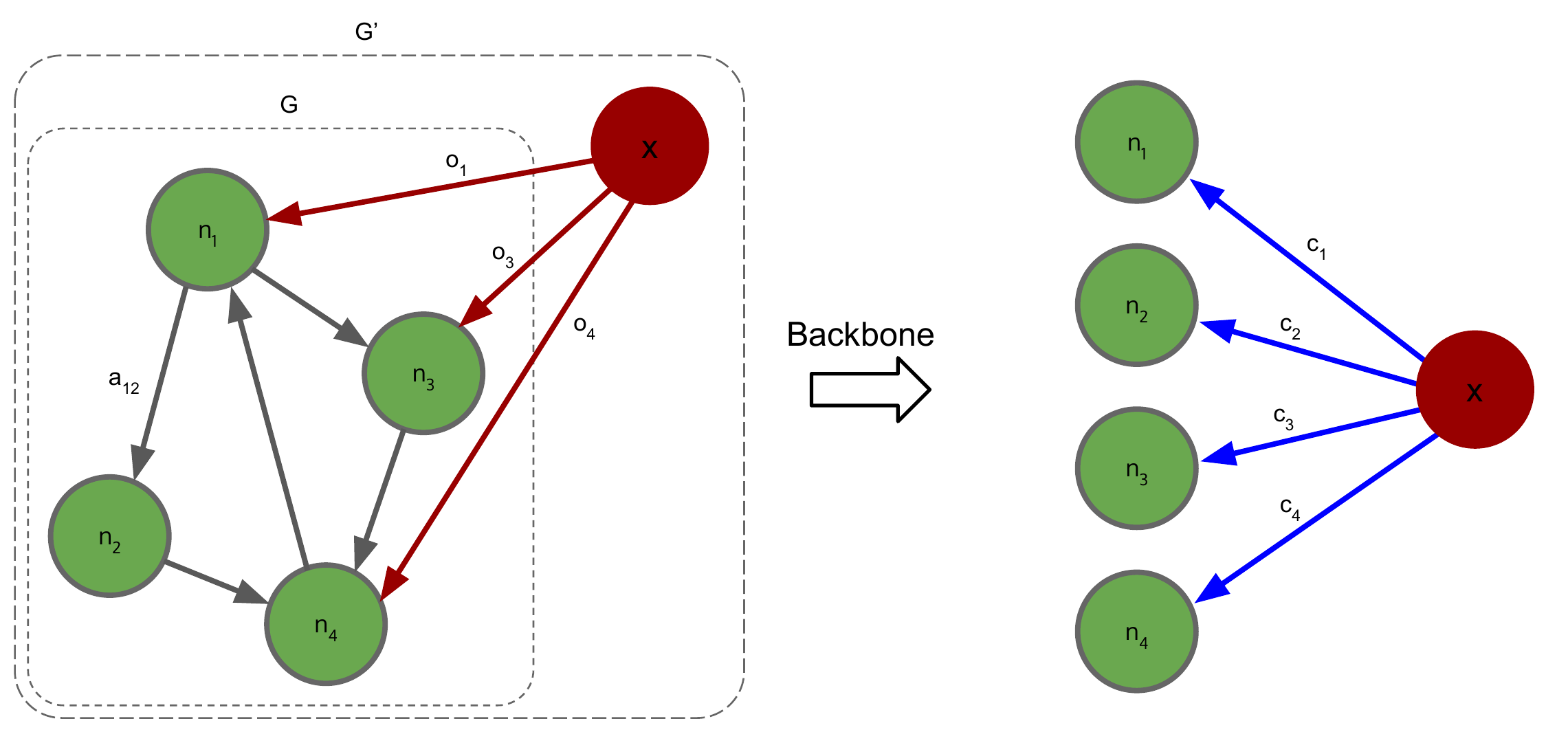}
    \caption{Example of a network $G$ with nodes $n_i \in \mathcal{V}$ (green) with values $v_i$, and edges $e_{ij}$ (grey) with their weights determined by the weighted adjacency matrix $A = \left[a_{ij}\right]$. We also show the extended network $G'$ with the external node $x$ (red), which is connected to the nodes of $G$ by the edges $e_{xj}$ (red) with weights $o_j$. Here, the directly controlled nodes are $\mathcal{S} \in \{n_1, n_3, n_4\}$. After applying the backbone algorithm to propagate control, we obtain the total indirect control $c_j$ (blue edges) of $x$ in all reachable nodes. Hence, even though $n_2 \notin \mathcal W$ is not directly controlled by $x$, it can be indirectly controlled, e.g. through $n_1$.} 
    \label{fig:network}
\end{figure}

We are interested in optimizing the control over nodes (individuals, companies, etc.) in the network. For this reason, we focus on how and to what extent a single external agent (or node), representing, for instance, a hostile company, can take control of a network. 
Consider a network $G$ of $N$ nodes $n_i \in \mathcal{V}$ with corresponding node values $v_i$, and (weighted) edges $e_{ij} \in \mathcal{E}$. The edges $\mathcal{E}$ between the nodes in $\mathcal{V}$ are represented by the weighted adjacency matrix $A = \left[a_{ij}\right]$, with $a_{ij}$ the weight of the edge $e_{ij}$ from node $n_i$ to node $n_j$. To represent the agent (whose intervention on the network we aim to optimize), we add an additional node $x$ to the network, which is attached to $G$ through a set of edges $e_{xi} \in \mathcal{O}$ that point to a selection of nodes $\mathcal{S} \subseteq \mathcal{V}$. $G'$ denotes the extended graph with nodes $\mathcal{V} \cup \{x\}$ and edges $\mathcal{E} \cup \mathcal{O}$. A schematic example is shown in Fig.~\ref{fig:network}. To differentiate between edges within $G$ and edges originating from $x$, we use edge weight $o_{j}$ to represent the direct control of $x$ on node $n_j \in \mathcal{S}$, with $\vec{o} = [o_1, ..., o_N]^T \in [0, 1]^{\abs{S}}$. We assume the existence of a backbone algorithm (BB) that can propagate the (indirect) control through the network. Propagation of control refers to the fact that the control of node $n_i$ over node $n_j$ also indirectly introduces control of $n_i$ over the children of $n_j$, and more generally, all of its descendants. The latter, however, requires a careful definition of how control is computed, e.g.\ in the case of cycles~\cite{vitali2011network}, and depends strongly on the chosen BB.
The result of the BB must be a vector $\vec{c} = [c_1, ..., c_N]^T \in [0, 1]^N$ that determines the total control (direct+indirect) of $x$ into all (reachable) nodes in $\mathcal{T}$ (see Fig.~\ref{fig:network}). In general, a node $n_j$ is reachable from another node $n_i$ if there is a directed path from $n_i$ to $n_j$. For unreachable nodes $c_j=0$. The BB algorithm must be continuous and differentiable in the edge weights $o_{j}$. Any method that meets the mentioned requirements (continuous, differentiable and producing a control vector) can be used as a backbone in our method.

Our goal is to optimize the (indirect) control over a targeted set of nodes $\mathcal{T} \subseteq \mathcal{V}$, by optimizing the direct control weights $o_j$ over a subset of nodes $\mathcal{S} \subseteq \mathcal{V}$. Note that when $\mathcal{S} \subset \mathcal{V}$ or $\mathcal{T} \subset \mathcal{V}$ are strict subsets, we use the terms \emph{source} and \emph{target} restricted control. Gaining control in node $n_j$ also comes with a cost, which is determined by the associated node's value $v_j$. The trade-off between cost and control can be tuned with the parameter $\lambda$, which effectively constrains the available budget. Hence, our optimization boils down to finding $\vec{o}^*$ for which

\begin{align}
    \vec{o}^* = \argmin{\vec{o}} \mathcal{L}_{control}(\vec{o}) + \lambda \mathcal{L}_{cost}(\vec{o}) .
    \label{eq:total_loss}
\end{align}

As control-loss function, we define (assuming $c_j\in [0,  1]$)

\begin{align}
    \mathcal{L}_{control}(\vec{o}) = \sum_{n_j \in \mathcal{T}} 1 - c_{j}(\vec{o}) \label{eq:Lcontrol}
\end{align}

which describes the \emph{lack} of control. Since we minimize the loss, this maximizes the control of $x$ over $\mathcal{T}$. For most use cases the complexity and non-linearity are dictated by the functional dependence of $c_{j}(\vec{o})$. For example, as we will show in the next section for the case of corporate control, the intervention of an external agent (given by $\vec{o}$) will change the adjacency matrix $A$, i.e.\ $A(\vec{o})$, which in turn affects the control: $c_{j}(A(\vec{o}), \vec{o})$.
The details of the loss function depend strongly on the use case. One may, for example, simply impose a cost that penalizes the number of takeovers

\begin{align}
    \mathcal{L}_{cost}(\vec{o}) =  \Lp{\vec{o}}_p
\end{align}

where $p=1, 2$ and $\Lp{}_p$ indicates the $l_p$-norm, thereby inducing sparseness in $\vec{o}$. To relate the problem to SIM and (exact) controllability: this optimization aims at maximizing the control, with as few takeovers as possible, while ignoring the node values. By defining a cutoff $C_{cut}$ for $c_i$ or using proximal gradient descent, we can obtain a proxy for the size of the typical control set. 
Another useful loss function may be

\begin{align}
    \mathcal{L}_{cost}(\vec{o}) =  \sum_{n_j \in \mathcal{S}} o_j v_j \label{eq:Lcost}
\end{align}

which represents the cost of the direct control in the set $\mathcal{S}$, weighted by the node values.

We have explicitly introduced an external agent for clarity. In practice, this node might coincide with an internal node, be more connected, or contain an alliance of nodes. Our framework, however, is generic, and can be easily extended to these specific cases. While $\lambda$ effectively constrains the budget, it might not be evident from its value what the budget exactly is. To accommodate situations where the optimisation needs to be run for a specific monetary value for the budget, we also implemented a constrained optimisation version of our framework (see Methods).


\section*{Results}
To show the power of our optimal control framework, we apply it to the corporate control problem and demonstrate its ability to provide actionable insights which were hitherto intractable to calculate. 

\subsection*{The Global Corporate Ownership Network}
In the context of global corporate control, information on the corporate ownership and corporate financials is typically available from government mandated financial reports that must be filed periodically. Using such data, Vitali \etal~\cite{vitali2011network} have designed an algorithm to calculate the global control of companies by propagating and consolidating direct and indirect ownership throughout the ownership network. Their algorithm essentially assumes a multiplicative process of control along the paths in the networks 
and allows one to compute the control distribution of a snapshot of a network. It does not, however, provide an answer to governance questions relevant to investment screening mechanisms such as ``How vulnerable is this sector to takeovers?''. Answering such questions requires inferring the optimal perturbation of the edge weights in the network, which can be achieved with our methodological framework.

In the context of corporate control, the edge weights $a_{ij}$ represent the direct control of company $n_i$ over company $n_j$, while the node value $v_i$ represents the total value of company $n_i$. In the applications that follow, we use the algorithm in Vitali \etal~\cite{vitali2011network} (see Methods for technical details) as the BB in our framework to compute control in corporate ownership networks


Given a trial external control/ownership vector $\vec{o}$, the network properties are affected by this external ownership. We therefore adjust the ownership matrix $A$ using the following Assumptions: 
\begin{enumerate}
    \item $x$ obtains a fraction $o_j$ of ownership in a company $n_j$ by uniformly buying stock from all companies $n_i$ that have ownership in $n_j$: $a_{ij} \to (1-o_j) a_{ij}$ (maximum entropy method). We define $A(\vec{o})$ as the adjusted ownership matrix for a given $\vec{o}$. To obtain ownership in root nodes of $G$, $A$ remains unaffected.\label{item:assumption_max_entr}
    \item we assume that $x$ can only obtain as much ownership in a node $n_j$ as explained by $G$: $o_{j}^{max} = \sum_{n_i \in \mathcal{V}} a_{ij}$ (except for root nodes of $G$, where we artificially set $o_{j}^{max} = 1$ in our examples).\label{item:assumption_max_stock}
\end{enumerate}
Assumption 1 introduces non-linearities in our external agent optimization problem, since we introduce $\vec{o}$-dependence into the non-linear BB algorithm by adjusting the adjacency matrix ($B$ in~\eqref{eq:vitali}). In other words, non-linearities are introduced due to the fact that the agent's interventions affect the properties of the network which it is trying to control.

\begin{figure}[tb]
    \centering
    \includegraphics[width=0.6\textwidth]{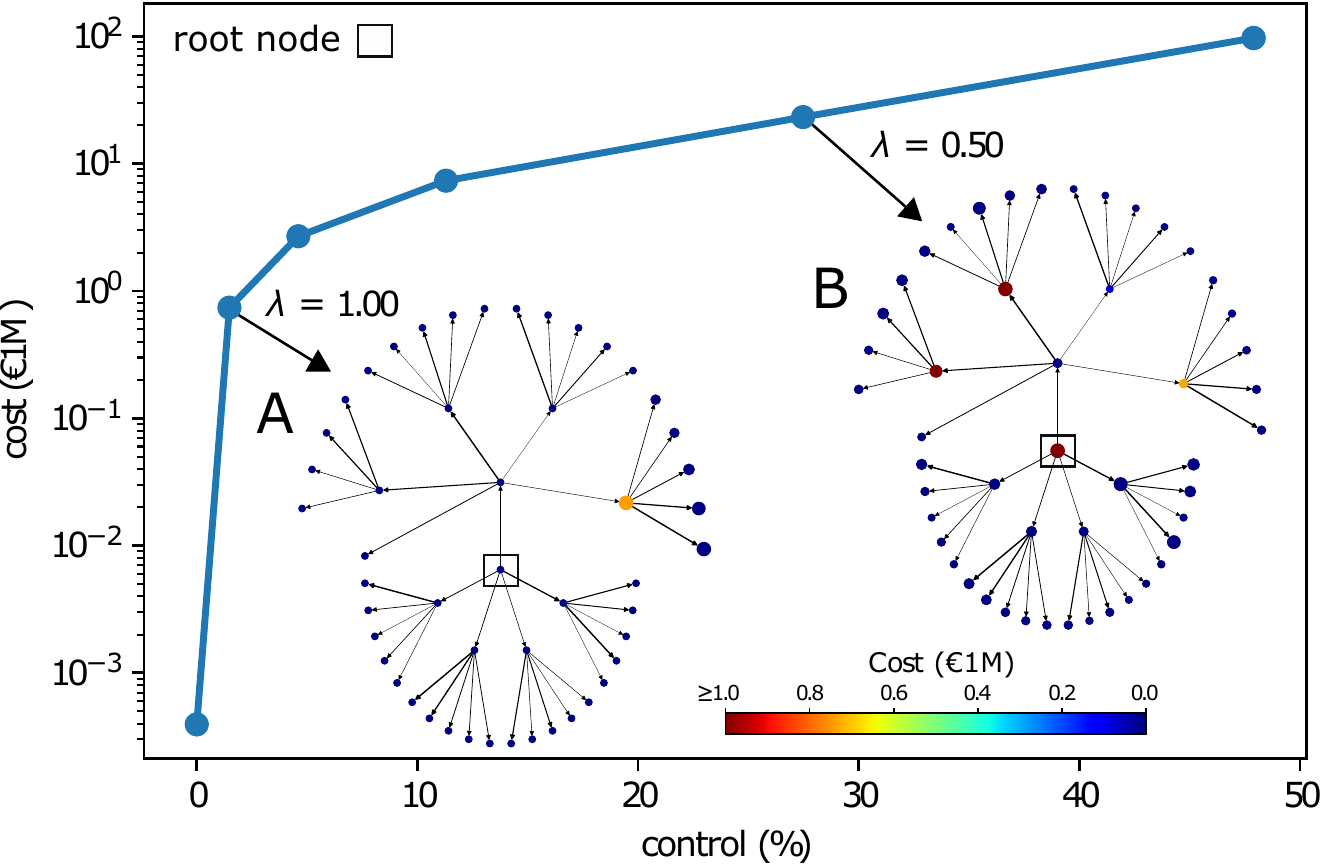}
    \caption{Control and cost loss as a function of $\lambda$ for the synthetic extended star graph. Control shows the \% of total shares of the target nodes that the external agent controls. Cost is the monetary value that the agent spends to acquire a given control (by buying source nodes). The insets \textbf{A}, and \textbf{B} show the intervention strategy of the external agent for $\lambda=1$ and $\lambda=0.5$ (in units of control per €1M). The color of the node represents the cost the external agent invests in that specific company, while the size represents the control the external agent acquires.}
    \label{fig:lambda_curve}
\end{figure}

\subsection*{Application to a synthetic graph} 
We first walk through the simple synthetic example of an extended star graph. This kind of topology appears frequently in corporate networks due to mergers and acquisitions between star networks consisting of companies and their subsidiaries. An example of this is Microsoft Corp, which owns Xbox, Skype, Linkedin, which in turn own other subsidiaries. For details see Figure 1 in Rungi \etal~\cite{Rungi2017}. The ownership weights are sampled from a uniform distribution $\mathcal{U}(0,1)$. The node values follow the simple rule that the central nodes (root nodes) have the highest value, while the value of other nodes declines with their distance from these root nodes. We implement this feature using $v_i = 2^{D-d_i+1}$ (in €1M), where $d_i$ and $D$ are the depth of node $n_i$ and the star (tree) respectively.

To study the vulnerability of the network to external control, we optimize Eq.~\ref{eq:total_loss} for several values of $\lambda$,
the results are shown in Fig.~\ref{fig:lambda_curve}. The $\lambda$-curve in Fig.~\ref{fig:lambda_curve} demonstrates exactly how expensive it is to obtain control over the network. Its changing slope, which reflects the changing trade-off between cost vs control, gives us an indication of vulnerability or susceptibility of the network to external control (e.g. hostile takeovers). We visualise the strategy of the external agent along the curve in insets \textbf{A} and \textbf{B} for $\lambda=1$ and $0.5$ respectively. The external agent prefers to take over companies closer to the root node up to the point where the budget is large enough to take over the root node itself, resulting in maximum control over the network. Note that not necessarily all shares of every company are available in the network (e.g. restricted shares). For this reason, the upper bound for external control might be less than $100\%$. In this network the preference for proximity to the root node reflects that nodes closer to the root node offer a more significant amount of indirect control (see Fig. S1). We can conclude that for the extended star graph it quickly becomes very expensive to gain control over the network. However, once the budget allows to gain control over the root node, maximum control is quickly reached. More detailed investigation on the exact distribution of costs over the nodes is possible with our methodological framework but falls outside the scope of this article. Still, we provide an example in Fig.S2 and Fig.S3 in the SI.

This simple example shows how the framework can be used to perform a detailed analysis of both network features such as susceptibility to control, and node-specific features such as the centrality of the nodes that offer maximal control. 

\subsection*{Application to the Great Britain (GB) biotech research sector}

To demonstrate the method in a real-world setting, we study GB's biotech sector. We extract the corporate ownership network from the Orbis Europe database provided by Bureau van Dijk, a private publisher of business information. We retrieve all the biotechnology research centers registered in GB and their in-component from the 2017 ownership snapshot of the European corporate network (See Methods for details). For simplicity we focus on the largest connected component. This network consists of $1109$ nodes and $1506$ edges, of which $114$ biotechnology research companies from GB. We take the biotech research centers as the target nodes ($\mathcal{T}$) over which we want to maximize control with a given budget. We simulate different restrictions on the source nodes ($\mathcal{S}$), reflecting different GB policies  with respect to foreign takeovers, to evaluate these policies' impact on the vulnerability of the biotechnology sector. We study three policy options:

\begin{enumerate}
    \item GB constraint: the external agent cannot buy a company registered in GB directly but can buy any other European company in the network. In this case the nodes in $\mathcal{S}$ contain \emph{no GB} companies. 
    \item GB biotech constraint: the external agent cannot buy a biotech research center registered in GB directly but can buy any other European or GB company in the network. In this case the nodes in $\mathcal{S}$ contain \emph{no GB biotech} research centres.
    \item Unconstrained: the external agent can buy any company in the network. In this case the nodes in $\mathcal{S}$ contain \emph{all} companies.

\end{enumerate}

\begin{figure}[tb]
    \centering
    \includegraphics[width=0.6\textwidth]{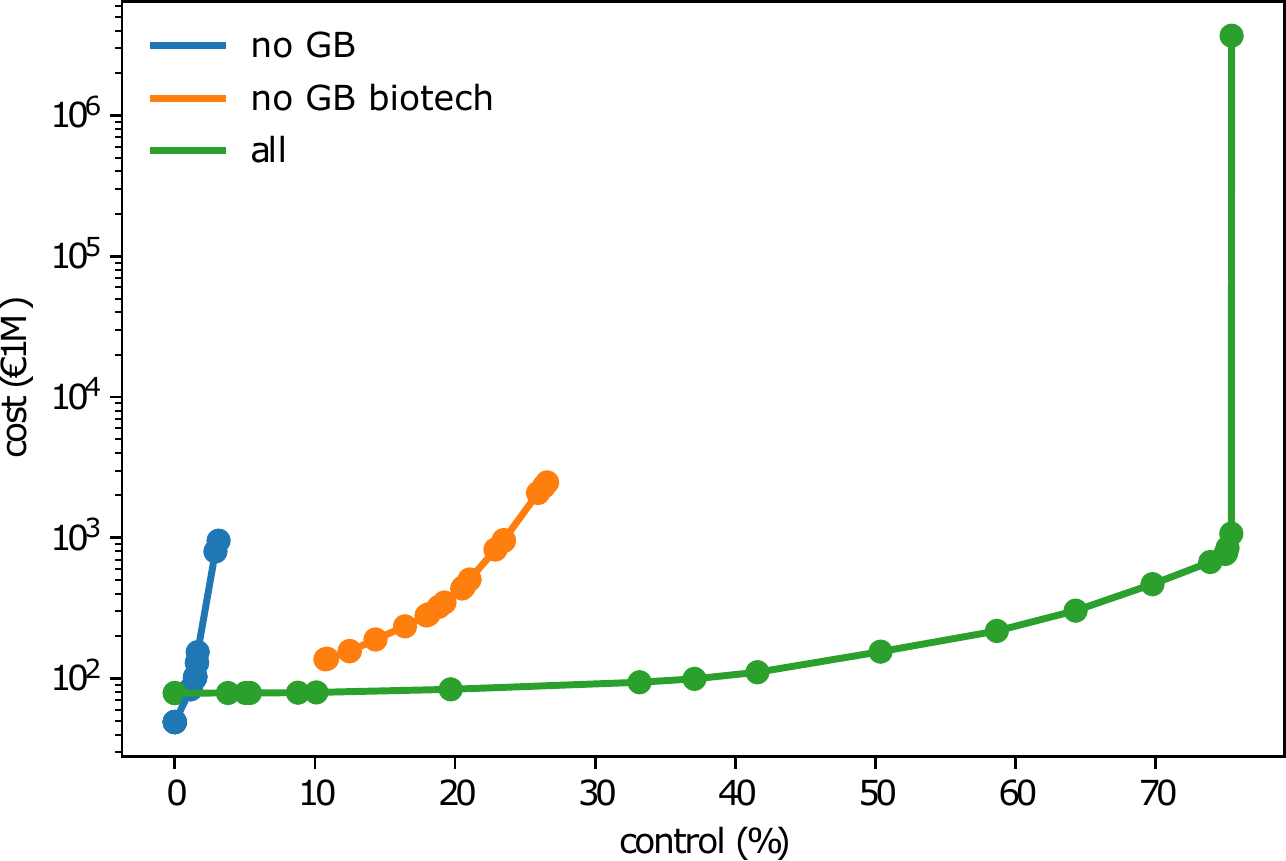}
    \caption{Control and cost as a function of $\lambda$ for the target restrictions: GB constraint (\emph{no GB}, in orange), for the GB biotech constraint (\emph{no GB biotech}, in blue), and for the unconstrained case (\emph{all}, in green). Control shows the \% of total shares of the target nodes that the external agent controls. Note that not necessarily all shares of every company are available in the network (e.g. restricted shares). Cost is the monetary value that the agent spends to acquire a given control (in €1M).}
    \label{fig:biotech}
\end{figure}

We first show the $\lambda$-curves for these policy environments in Fig.~\ref{fig:biotech}. The unconstrained case can be viewed as an upper limit. Here the external agent is allowed to take over all available shares of the target companies directly, provided sufficient budget is available. From this unconstrained case, we see that just below $80$\% represents the upper bound for control in the GB biotech research centers. Because buying direct control in the biotech research centers is not allowed in the GB and GB biotech constraint cases, control over the target companies can only be obtained indirectly through other companies. The optimal strategy to gain control over the target set in these cases is thus to leverage the ownership network and consolidate indirect ownership into control, a strategy that is often not accounted for in policy because of the complexity of ownership networks. Even for the relatively small network surrounding the biotech research centres, indirect control is non-trivial to account for (see Fig.S2 for a snapshot of the network). Being able to do so highlights one of the main benefits of our optimal control framework. 

\begin{figure}[tb]
    \centering
    \includegraphics[width=0.6\textwidth]{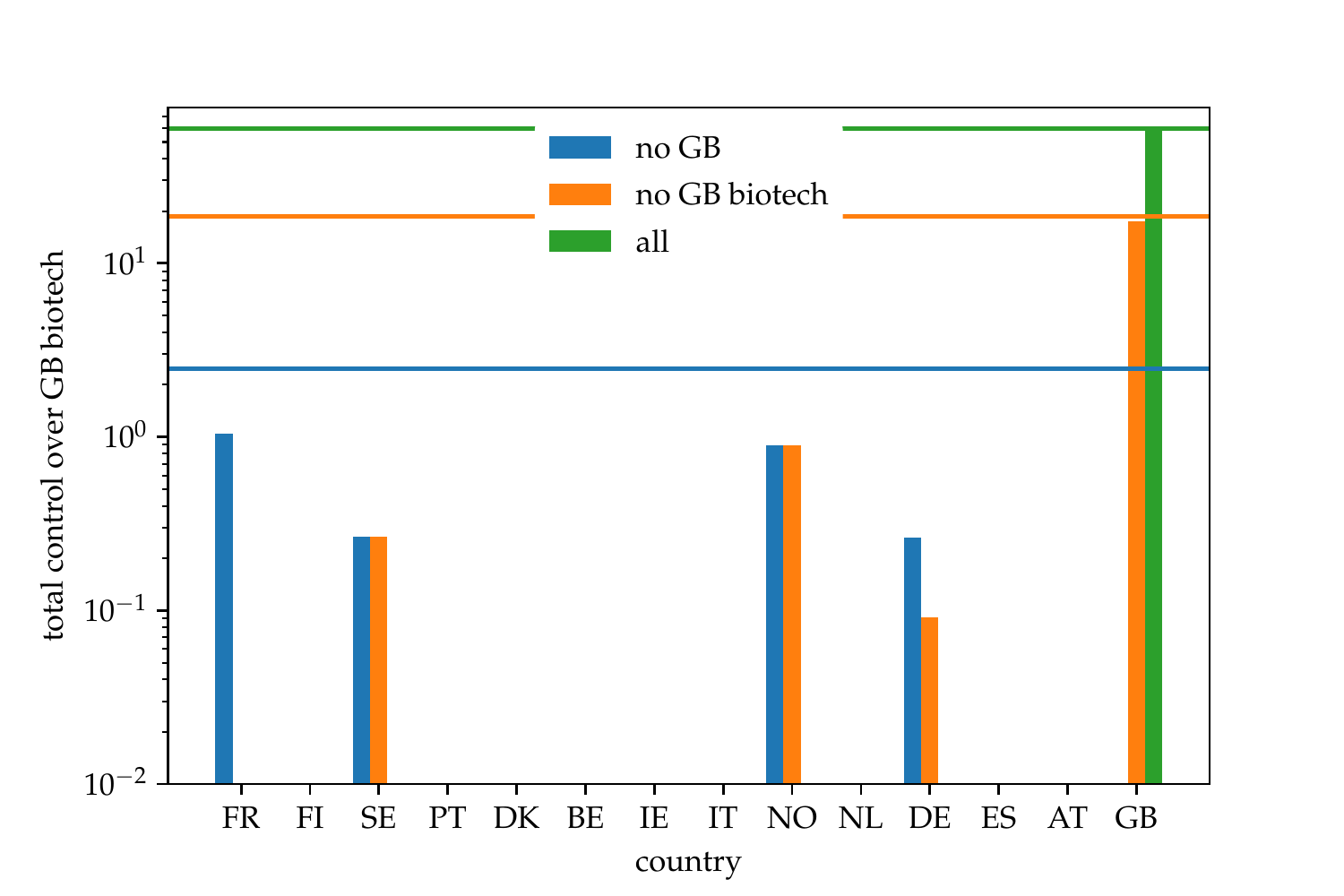}
    \caption{Distribution of the source of control over the various countries for a budget-restricted optimization. The budget is chosen to be $B=$€$901$M, which equals the total value of the target set of GB-biotech firms. The horizontal line indicate the total control achieved in the different policy environments.}
    \label{fig:country_distr}
\end{figure}

For the GB biotech constraint, i.e. GB specifically protects their biotech research centers from direct foreign investments, we find that a foreign investor can still obtain up to $30\%$ control over the sector. Most of this indirect control can be realized through non-biotech companies in GB (see Fig.\ref{fig:country_distr}), who thus seem to be more closely connected to the biotech research centres than (biotech) companies in the EU. If the constraint is  extended to restrict investments in all GB companies (GB constraint), we find that only $4\%$ of the sector is exposed to foreign control through EU companies. Note that these exposures represent lower bounds since we currently do not allow the external investor to change the outgoing edge weights (ownership) of the companies it invests in. 

To fully protect the target set from unwanted external control, it is thus clear that policy should take into account the complete network of ownership and especially the large indirect exposure of the network.

\section*{Discussion}
Controlling the many socioeconomic networks that greatly influence our daily lives has been a longstanding challenge in network science and across many fields. To go from observing and describing towards governing, there is a need to calculate optimal interventions. Inspired by building blocks from network controllability and social influence analysis, we reformulate the problem of optimal interventions in networks into an optimization of an external agent's objective function. Herein, a first term captures the control objective, and the second term the cost of the control. This objective function is subsequently optimized using automatic differentiation and gradient-based optimization strategies. To the best of our knowledge, our framework represents the first occasion where automatic differentiation and gradient-based optimisation are integrated into a coherent framework for interventions in the context of (social) networks. Probably closest to our work are Fan \etal~\cite{Fan2020} who use deep reinforcement learning to find key nodes in tasks such as network destruction. While they search for optimal strategies to alter network connectivity (network property), our work allows to optimize a property on the network (node properties). We also point out that in machine learning, a similar reformulation of learning causal interactions as a continuous optimization problem gave rise to an entirely new field of causal deep learning and causal representation learning~\cite{zheng2018dags}.

By building on a well-established backbone algorithm to calculate corporate control~\cite{vitali2011network}, we show that our framework is capable of characterizing the vulnerability of corporate networks in strategic sectors to sensitive takeovers, an important policy challenge for many countries (e.g. the EU framework for screening foreign direct investment started October 2020). Insights derived from our method could facilitate the protection of strategic sectors from unwanted foreign control.
Though we restricted the demonstration of our framework to the specific context of corporate control, it is directly applicable with the same BB algorithm by Vitali~\etal~\cite{vitali2011network} to a wide class of problems that require maximizing the diffusion of control, information, or influence throughout a network. This is because Vitali~\etal~\cite{vitali2011network} is valid in any context with a multiplicative process along the edges of a network (e.g. probability-based propagation). For example, re-imagining the synthetic graph in Figure~\ref{fig:lambda_curve} as a specific socioeconomic network, the insights can be reinterpreted for use in a marketing or information strategy. Further, any other propagation algorithm that is continuous and differentiable can be leveraged in the proposed framework, making it flexible to different use cases where propagation might not be multiplicative. 

Our research shows that combining state-of-the-art gradient-based optimisation techniques and differentiation tools (that are mainly used in deep learning) with network theory opens up new opportunities for complex systems research to better understand, design, and govern networks. The present work represents an important first step in exploring these opportunities. In future work, we plan to extend the framework to calculate optimal interventions in a multi-agent context, where the different agents can have competing objectives in a shared environment. This will further broaden the applicability of the framework to more complicated scenarios, like in the case of foreign entities and the government as two separate agents competing for control. We expect that in the future, our method can be used to increasingly larger networks, where the current development of tools (such as PyTorch) for sparse matrices is of primary importance.

\section*{Methods}
\subsection*{Vitali backbone algorithm for corporate control}\label{sec:vitali}

As an example backbone algorithm to compute the indirect control in corporate networks, we use the corporate control algorithm by Vitali \etal~\cite{vitali2011network}. We briefly outline their methodology. We first discuss how to compute corporate control between any pair of nodes $n_j, n_i \in \mathcal{V}$ within the network $G$ in Fig.~\ref{fig:network}, and then show how to obtain the control of the external agent.

Corporate control is approximated through corporate ownership. We assume we have a corporate ownership network, where the adjacency matrix $A = [a_{ij}]$ quantifies the factional ownership of company or node $n_i$ in company $n_j$. When company $n_k$ owns a fraction $a_{ki}$ of company $n_i$, it is common to assume in the corporate control literature that $n_k$ indirectly owns $a_{ki}a_{ij}$ shares of firm $n_j$ through the path $n_k \to n_i \to n_j$.

We introduce the matrix $B$, which is obtained by selecting the rows and columns from (the direct control matrix) $A$ that correspond to nodes that are reachable from $n_i$ (removing incoming links in $n_i$) with path length $1$. These are simply the children of $n_i$. The corresponding control of $n_i$ into any $n_j$ is given directly by $C_{l=1} = [B]_{ij}$. Next, we consider all directed paths of length $2$ starting from $n_i$. Through these paths, $n_i$ obtains an additional control over a set of companies that are reachable with path length $2$, where the corresponding control of $n_i$ in $n_j$ is given by $C_{l=2} = [B^2]_{ij}$. Continuing this way, and taking into account all possible path lengths, we find that the propagated control $c_{ij}$ of any node $n_i \in \mathcal{V}$ in any other node $n_j \in V$ via the network $G$ is given by the matrix  $C = [c_{ij}]$ as

\begin{align}\label{eq:vitali}
    C = \sum_{l=0}^{+\infty} C_{l} = (\iden{} - B)^{-1} 
\end{align}

where $C_{l=0} = \iden{}$. For a more in-depth discussion, we refer to Vitali \etal~\cite{vitali2011network}.

By attaching the external party $x$ to node $n_i \in \mathcal{S}$ with an edge $o_{j}$, $x$ obtains control in a company $v_j$ through all paths connecting $n_i$ to $n_j$. The control $C_{ij}$ is  weighted by the edge control $o_j$ of $x$ in $n_i$. Hence, the total control of $x$ over nodes $\mathcal{V}$ is given by

\begin{align}
    \vec{c}^T = \vec{o}^T(\iden{} - B)^{-1} 
\end{align}

One main difficulty in strongly connected corporate control networks is the presence of cycles, which the method of Vitali \etal~\cite{vitali2011network} solves correctly. Nodes that are not reachable from $x$ are of course not controllable by $x$: $c_j = 0$. We point out that the form of ~\eqref{eq:vitali} can be used in a more general context than corporate control, since it accumulates the total signal by tracing all paths of all lengths between two nodes in the network.

\subsection*{Core Method}
In general, we take a network $G$, with node set $\mathcal{V}$ and edge set $\mathcal{E}$, to which we associate a set of (continuous-valued) node and edge property vectors: $V = \{v_i | n_i \in \mathcal{V}\}$ and $A = \{a_{ij} | e_{ij} \in \mathcal{E}\}$. In the specific case of node control in corporate networks, $v_i \in \mathbb{R}$ is the company value and $a_{ij} \in \mathbb{R}$ is fraction of company $n_j$ that is owned by company $n_i$. We point out, however, that both $v_i$ and $e_{ij}$ can be vectors. Furthermore, their values can be fixed for some $i,j$. In our case of node control with an external agent, only a subset of the $e_{ij}$ is varied directly, namely those that are outgoing of the external agent. The other $e_{ij}$ are adjusted deterministically based on the latter, as discussed in Assumption~\ref{item:assumption_max_entr}. Depending on the application, we define a cost function with two terms. A first term is a control loss function $\mathcal{L}_{\textrm{control}}(V, E; G)$ representing the amount of control a set of nodes has over the network, for a given $V$ and $E$. A second term $\mathcal{L}_{\textrm{cost}}(V, E; G)$ penalizes domains of the phase space $(V,E)$ that carry a high cost. Given that the total loss function, $\mathcal{L}_{\textrm{control}} + \lambda \mathcal{L}_{\textrm{cost}}$ (with $\lambda$ a constant factor that balances the importance of control and cost) is continuous and differentiable (almost everywhere) in the node and/or edge properties that we aim to optimize for, we determine the optimum via automatic differentiation and gradient descent. 

In the above $\lambda$ quantifies importance of cost versus control and enables us to gain insights into the general cost to control the system (by varying $\lambda$). In many applications the optimisation will involve a monetary budget constraint instead of a relative importance measure. In this case, the total cost that corresponds to a given set of properties $V,E$ is subject to the constraint of a (maximum) budget $B$. We then need to optimize

\begin{align}
    \min_{\vec{o}} \mathcal{L}_{control}(V,E; G) \nonumber\\
    s.t. \quad H_{budget}(V,E; G) = 0
    \label{eq:constrained_loss}
\end{align}

where $H_{budget}(V,E; G) = \mathcal{L}_{cost}(V,E; G) - B$ (the above may also be an inequality constraint).
To solve the latter problem, we use the augmented Lagrangian approach~\cite{nemirovsky1999optimization, zheng2018dags}, which uses an unconstrained optimization problem (with a quadratic penalty for the constraint) to solve the constraint optimization task. To account for budget constraints, we can thus optimize the dual loss function with Langrange multiplier $\alpha$:

\begin{equation}\label{eq:augmented}
\begin{aligned}
    \mathcal{L}_{augm}(V,E; G) = \mathcal{L}_{control}(V,E; G)&
    +\frac{\rho}{2} H_{budget}(V,E; G)^2\\
    + \alpha \left| H_{budget}(V,E; G)\right|.
\end{aligned}
\end{equation}

\subsection*{Optimization procedure}
In practice, we parametrize the optimization problem with a set of unbounded variables $\vec{p}_u \in \mathbb{R}^M$ and use the sigmoid function ($\sigma$) to obtain an ownership fraction $\vec{p} = \sigma(\vec{p}_u) \in [0, 1]^M$. Furthermore, assumption \ref{item:assumption_max_stock} suggests that $o_j \in [0, o_{j}^{max}]$. Therefore, we parametrize $o_j = o_{j}^{max} p_j$ to remain within these bounds. Hence, $o_j$ represents to total fraction of company $n_j$ owned by $x$, while $p_j$ represents the fraction of the \emph{available} stock owned by $x$. By implementing the procedure in deep-learning toolboxes such as PyTorch, one can perform network optimization use-cases efficiently using autograd. More specifically, the gradients are computed through a reverse-mode gradient accumulation scheme~\cite{NEURIPS2019_9015}.
We use learning rates from the set $\{1, 0.1, 0.001\}$ and the Adam optimizer with default parameters in the PyTorch implementation. Notice that PyTorch also offers the possibility for sparse matrices, which would allow to scale the approach to large networks. The parameters $p_{uj}$ are initialized by generating random values from a normal distribution $N(\mu=-7, \sigma=10^{-4})$, or by estimating the values of $\vec{o}$ by the fraction of a company value with respect to the available total budget. The optimization stops when the loss changes by less than $10^{-8}$ in $5$ subsequent optimization steps, or after a maximum of $3k$ steps, whichever occurs first. As an example, a complete single optimization, for a given value of $\lambda$, takes about 10mins for the Orbis Europe dataset of 1048 nodes on a laptop with a 1,4 GHz Quad-Core Intel Core i5 processor.
For the constrained optimization, we optimize the augmented Lagrangian in \eqref{eq:augmented} with $\rho=1$ and $\alpha=0$. 
We iteratively increase the value of $\alpha$ by adding $\rho  |H_{budget}|$ and re-optimize. 
We then increase $\rho$ by scaling it with a factor of $10$, whenever we obtain a new value for the constraint $| H_{budget}|$ larger than $0.25$ times the last optimal value of $ |H_{budget}|$. This process is repeated until $ |H_{budget}|$ is below a given tolerance threshold. 

\subsection*{Orbis Europe biotech data}\label{sec:orbis_data}
The Orbis Europe database is compiled by Bureau van Dijk, a private publisher of business information who collects ownership structure and financial information on almost $20$ million companies in Europe. If a company outside of the EU owns part of a European company, it will also be included in the data, financial information will however not be available for this non-EU company. From this database we extract all European or British companies for which total assets are reported and take this as our base network. From this network, we select the companies registered in Great Britain with NACE-code (standard European industrial classification code) $7211$, whose activities are defined as: \textit{``Research and experimental development on biotechnology''}. This resulted in $260$ companies. Taking the in-component of all these companies (the companies that via some ownership path can reach any of the $260$ biotech companies) resulted in a network of $1381$ nodes and $1695$ edges. From this we only take nodes with a known NACE-code so that we only extract the corporate network and drop for instance the individual owners. For simplicity and plotting purposes, we further only keep the largest connected component in this network. This final selection resulted in a network of $1109$ edges and $1506$ nodes, with $114$ of them being biotech research centres in Great Britain.

\subsection*{Code availability}
The code of the Optimal Network Control framework and  for the reproduction of the figures is detailed and will be made available in the following GitHub repository: \url{https://github.com/CSI-ADS/OptimControl}
\subsection*{Data availability}
All data analysed in this paper, including synthetic graphs and real-world networks, are anonymised according to the conditions of the data provider (Bureau van Dijk) and included in the GitHub repository.

\section*{Author contributions}
J.N., M.v.d.H designed research; J.N., M.v.d.H. performed research; J.N. implemented the model; J.N., M.v.d.H. analyzed and processed results; J.N., M.v.d.H wrote the original draft; J.N., M.v.d.H, K.S, and B.M reviewed and edited the final draft; and K.S., B.M. provided data.

\clearpage
\bibliography{nature-bib.bib}

\newpage
\renewcommand\thefigure{S\arabic{figure}}
\setcounter{figure}{0}
\section*{Supplementary Figures}\label{sec:suppl}
In this section, we show figures that fall outside the scope of the main text but are valuable for the reader to gain insight into the flexibility and possibilities the proposed framework provides to study optimal control and the resulting distribution of control and cost in the network.
\subsection*{Synthetic extended star network}
\begin{figure}[htb]
    \centering
    \includegraphics[width=\textwidth]{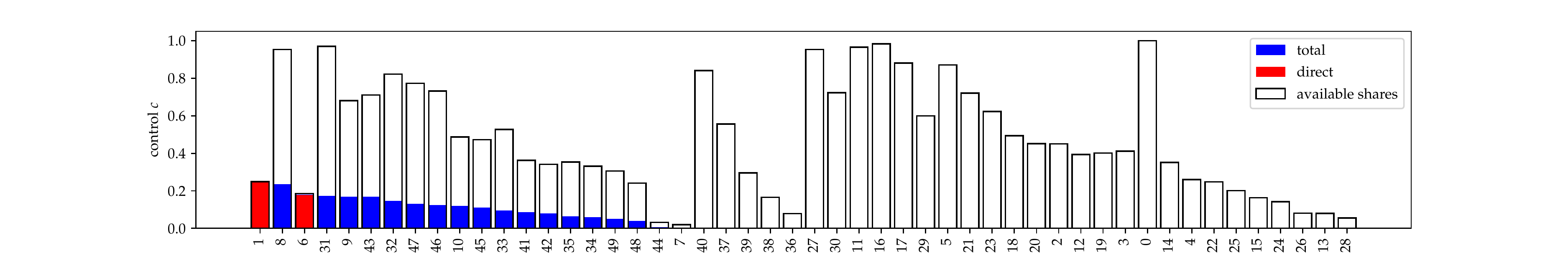}
    \includegraphics[width=\textwidth]{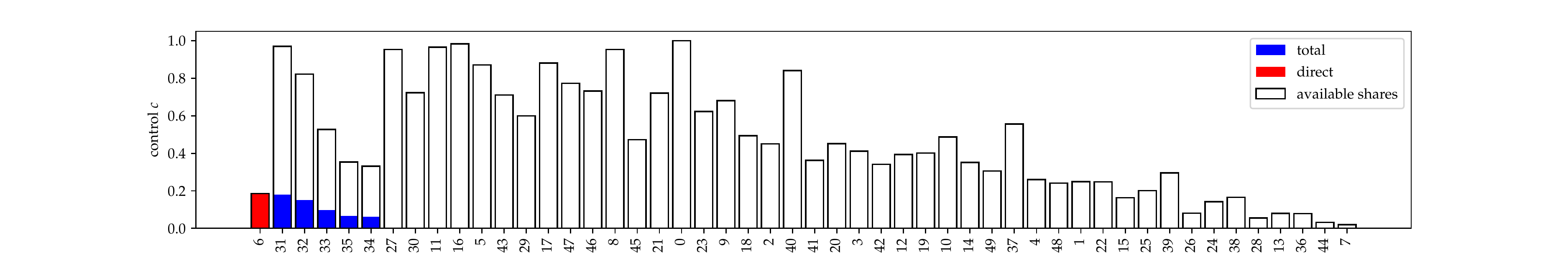}
    \caption{ Distribution of control gained at $\lambda=0.75$ (top) and $\lambda=1$ (bottom). We show the direct (red) and indirect(blue) control of the external agent for the star network. The companies on the x-axis are sorted in descending order of the total shares bought by the external agent. Total shares available are shown in black. We remind the reader that we artificially set the $o^{\textrm{max}}=1$ for root nodes (here company ``0'' on the x-axis).}
    \label{fig:control_distr_star}
\end{figure}

\begin{figure}[htb]
    \centering
    \includegraphics[width=\textwidth]{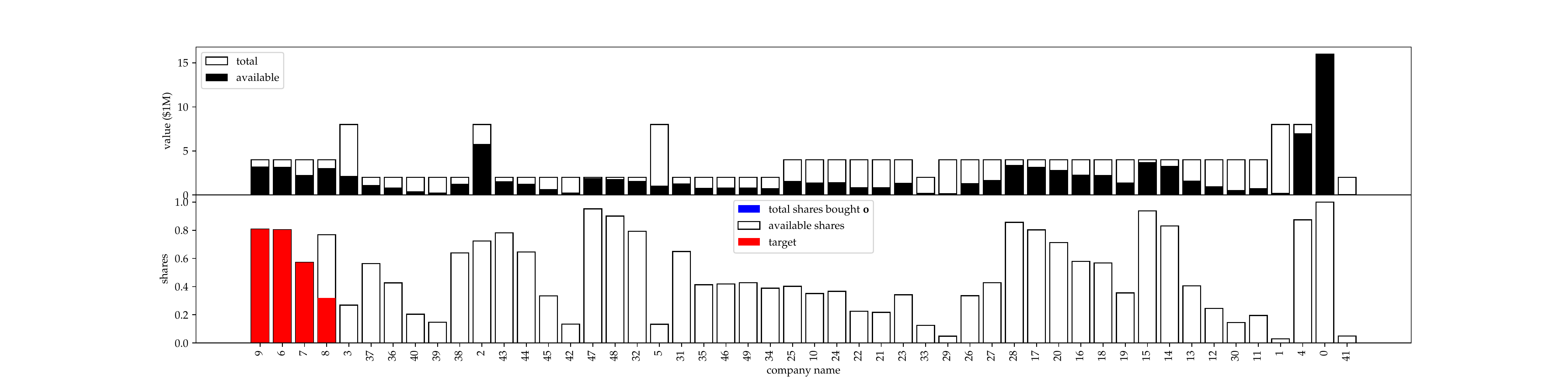}
    \caption{(top) Empty bars represent the total value of the companies in the star network, while the black bars indicate the fraction that is available in the network (determined by the sum of incoming weights). (bottom) Distribution of shares bought in the network by an external agent, where red bars reflect the fact that the considered company is also a target (which is the case for all nodes in the star network). The control distribution is obtained for a budget restriction of €10M.}
    \label{fig:star_o_distr_10}
\end{figure}

\begin{figure}[htb]
    \centering
    \includegraphics[width=\textwidth]{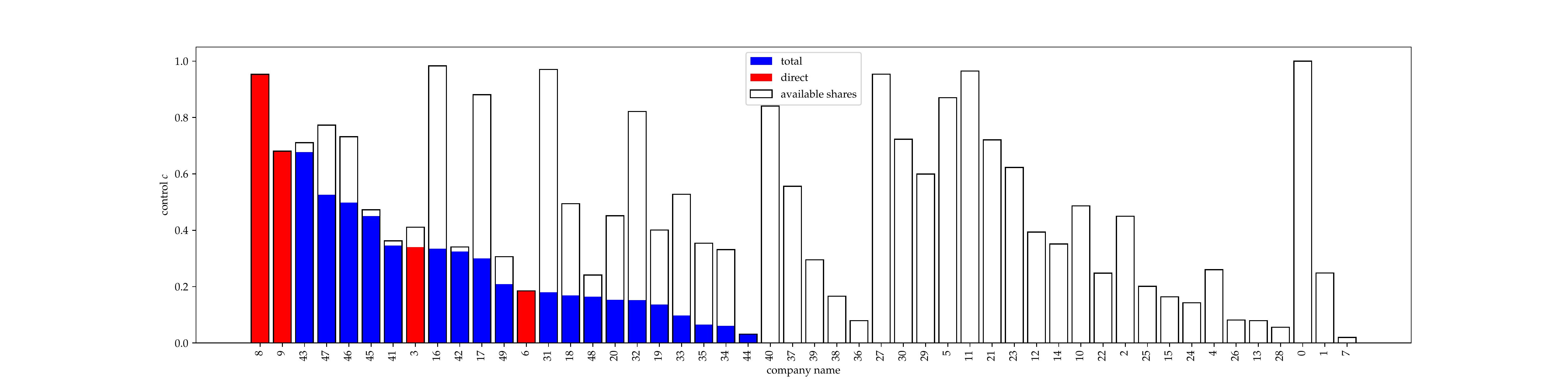}
    \caption{Distribution of control over the companies in the star network. The black bars indicate the total available shares for that company. Blue indicates the total control on a company, of which the fraction of direct control is indicated in red (often the entirety of the bar). The control distribution is obtained for the case in Fig.~\ref{fig:star_o_distr_10}.}
    \label{fig:star_control_distr_10}
\end{figure}

Fig~\ref{fig:control_distr_star}-\ref{fig:star_control_distr_10} shows more a more detailed view on the distribution of cost and control for the extended star-graph example in the main text. On a case-to-case basis, this can lead to more insights into how both direct and indirect control, and cost is distributed along the network and how it might be connected to other node-specific features (e.g. type of company, centrality).
\newpage

\newpage

\newpage

\clearpage
\subsection*{Great Britain biotech research centres}
\begin{figure}[htb]
    \centering
    \includegraphics[width=0.8\textwidth]{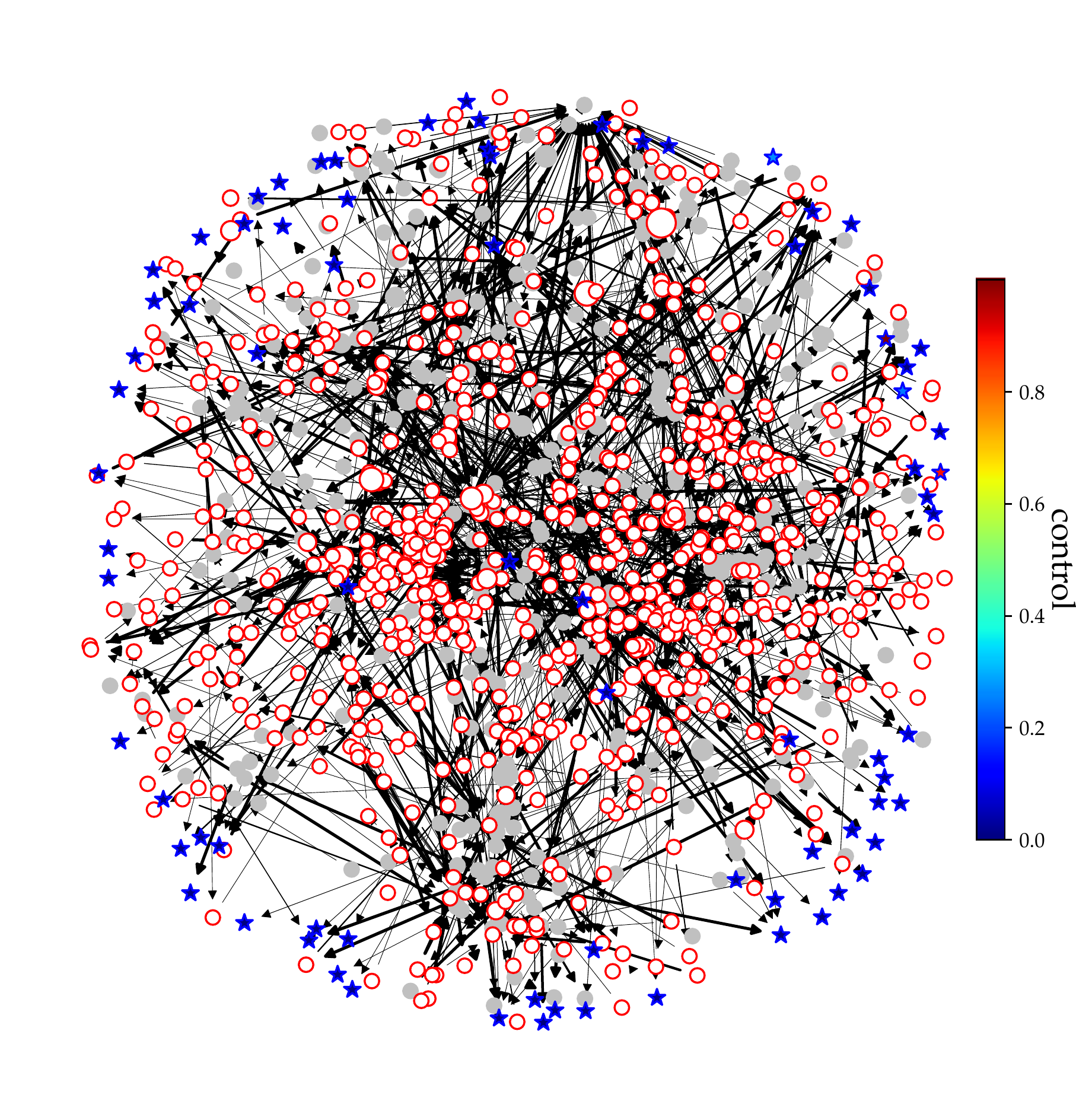}
    \caption{Amount of control (color) over biotech research centres (stars) in Great Britain as targets. The source nodes (red circles) and intermediary companies (grey) are shown for the country constraint policy environment (making all non-GB companies sources, represented by round nodes with red borders). The results are shown for the constrained optimization with a budget that equals total value of all targets. The size of the red circles reflects the cost of acquiring that company in the final result.}
    \label{fig:main_biotech}
\end{figure}

Fig~\ref{fig:main_biotech} shows the complexity of the relatively small network constructed from the largest connected component of the in-component of the biotech research centres in Great Britain. The obvious complexity of the network shows that automated techniques are needed to gain insights into the structure's effect on control in the network.

\end{document}